\documentstyle[12pt]{article}

\newcommand{\beq}{\begin{equation}}
\newcommand{\eeq}{\end{equation}}
\newcommand{\eq}[1]{eq. (\ref{#1})}

\title{
Radiative Corrections to the Light Muonic Atoms Decay Rate
}

\author{
Savely G. Karshenboim${}^{a,}$\thanks{
E-mail: sgk@onti.vniim.spb.su;
karshenboim@phim.niif.spb.su}
~and Vladimir G. Ivanov${}^b$,\\
\bigskip
{}\\
${}^a$ D. I.  Mendeleev Institute for Metrology,\\
198005, St.~Petersburg, Russia\\
${}^b$ Pulkovo Observatory, 196140, St. Petersburg, Russia\\
{}\\
\bigskip
\bigskip
Short title: Light Muonic Atoms Decay\\
{}
}

\date{}

\begin{document}

\large

\maketitle

\newpage

\begin{abstract}
\large
Radiative corrections to the decay rate in the low-$Z$ hydrogen-like
muonic atoms are considered. The correction arises from the Uehling
potential and it has the relative order of $\alpha$. The numerical results
are reported for the $2p\to 1s$ transition for the atom with the nuclear
charge value up to $Z=10$.
\end{abstract}

\newpage

The work is devoted to a calculation of the radiative corrections to the
decay rate in the light muonic atoms. Those arise from the
electronic vacuum polarization effects and it is expected to be on the
percent level. The percents level has also been obtained measuring the
ratio of some lines intensity in several muonic atoms with not too high
value of the nuclear charge $Z$ (see for details Ref. \cite{Ort} and
references there).

The non-relativistic expressions for the decay rate and intensity are well
known:

\beq                  \label{Gdip}
\Gamma(\lambda'\to \lambda) =
\frac{4 \omega_{\lambda'\lambda}^3}{3}
\vert {\bf d}_{\lambda'\lambda }\vert^2
\eeq

\noindent
and

\beq                  \label{Idip}
I(\lambda'\to \lambda) =
\frac{4 \omega_{\lambda'\lambda}^4}{3}
\vert {\bf d}_{\lambda'\lambda}\vert^2.
\eeq

\noindent
where relativistic units in which $ \hbar  = c = 1$ and $\alpha=e^2 $ are
used. The leading one-loop radiative correction in the low-$Z$ muonic atoms
can be approximated adequately by only the Uehling potential term,
which corresponds to a non-relativistic electrostatic
potential\footnote{The situation is in contrast with that for the "usual"
(electronic) atom, where the main radiative contribution is due to the self
energy of an electron in the Coulomb field.}. Thus, the equations given
above are also valid for the radiative corrections, for which the energies
and the dipole matrix elements have to be calculated for a Schr\"odinger
particle in a combined potential

\beq
V({\bf r})=V_C({\bf r})+V_U({\bf r}),
\eeq

\noindent where the first term corresponds to the Coulomb field and the other
associates with the vacuum polarization effects.

Before calculating directly, the correction can be easy estimated. The
estimate may be based on the fact that the Uehling potential yields the
energy shifts $\Delta E_U(nl)$ close to those arisen from the delta-like
potential

\beq   \label{delV}
V_{\delta}({\bf r}) = A\,\delta({\bf r}).
\eeq

\noindent The unequalities

\[
\delta_1 =
\frac{\Delta E_U(1s) - 8 \Delta E_U(2s) }{
\Delta E_U(1s)}  < 1
,\]

\noindent and

\[
\delta_2 =
\left\vert
\frac{\Delta E_U(2p) }{
\Delta E_U(2s)}
\right\vert  < 1
,
\]

\noindent
for the Uehling potential (see Table 1) have to be compared with the
equation

\[
\delta_{1,2} = 0
\]

\noindent
for the delta-like potential.

In case of the potential of \eq{delV} the decay rate correction is known
(see Ref. \cite{JETP94} for the $2p\to 1s$ transition and
Ref. \cite{IK1} for the transitions between some higher states
with $n\leq4$). The results for the $2p\to 1s$ decay are

\beq  \label{gdfcor}
\Delta \Gamma_\delta \simeq -0.98\,R_{\delta}\, \Gamma^{(0)}
\eeq

\noindent and

\beq  \label{idfcor}
\Delta I_\delta \simeq -2.32\,R_{\delta}\, I^{(0)}
,\eeq

\noindent
where the unperturbated values of $\Gamma^{(0)}$ and $I^{(0)}$ are
determined from \eq{Gdip} and \eq{Idip} within the pure Coulomb field, the
effective parameter

\[
R_{\delta} = \frac{\Delta E_{1s}}{E_0}
,\]

\noindent is expressed in terms of the energy shift of the ground state
by the potential and an analog of the Rydberg constant

\[
E_0=\frac{m_R (Z\alpha)^2}{2}
\]

\noindent
including the reduced mass of the muonic atom $m_R$.

The results for the $2p\to 1s$ transition of the estimate in the nuclear
charge range $1-10$ are summarized in Table 1. The final corrections
for the delta-like potential of \eq{delV} presented in Table 2 are results
of some cancellation of the $E$-contributions (corrections to the frequency
$\omega$ in \eq{Gdip} and \eq{Idip}) and $\Psi$-contributions (corrections
to the dipole matrix element). Hence, the uncertainties of the estimates in
Table 1 may be higher than na\"{\i}ve expectation on the level of
$max\{|\delta_1|,\delta_2\}$.

For higher value of $Z$ another reason may be used for estimating. The
non-relativistic decay rate and the intensity are proportional to
$(Z\alpha)^4$ and to $(Z\alpha)^6$, respectively. The vacuum polarization
effects can be also expressed in terms of the running coupling constant

\beq
Z \alpha (q) = Z \alpha \cdot \left\{ 1+\frac{\alpha}{\pi}
\left(\frac{1}{3}\log{\frac{q^2}{m_e^2}}-\frac{5}{9}\right)\right\}
,\eeq

\noindent
and for the muonic atomic states the characteristic value

\[
q(nl)=\frac{Z\alpha m_R}{n}
\]

\noindent
is to be used for calculating. The corrections to the rate and the
intensity evaluated according their scaling in $(Z\alpha)$

\beq
\frac{\Delta\Gamma_{\alpha}}{\alpha\Gamma^{(0)}}=
4\times\frac{1}{3\pi}\left(\log{\frac{(Z\alpha m_R)^2}{m_e^2}}
-\frac{5}{3}\right)
,\eeq

\noindent
and

\beq
\frac{\Delta I_{\alpha}}{\alpha I^{(0)}}=6\times\frac{1}{3\pi}\left(
\log{\frac{(Z\alpha m_R)^2}{m_e^2}}-\frac{5}{3}\right)
\eeq

\noindent
are contained in Table 3. The uncertainty arisen from the discrepancy
between $q(1s)$ and $q(2p)$ is assumed to be on the level of

\[
\delta_\alpha =  \frac{\log{2^2}}{
\log{\frac{(Z\alpha m_R)^2}{m_e^2}}-\frac{5}{3}}.
\]

\noindent
The estimate based on value of $q(1s)$ should be rather higher than the
exact value.

The results of the straightforward calculation for the $2p\to 1s$
transition for the nuclear charge $Z$ up to 10 are summarized in Table 4.
Their evaluation is based on \eq{Gdip} and \eq{Idip} and the corrections
($\Delta\Gamma_U$ and $\Delta I_U$) are defined as the differences between
the Uehling potential results and the pure Coulomb ones. The main terms are
labeled by superscripts $s$ and $p$ associated with the state ($1s$ and
$2p$), which the correction is computed for, and by subscripts $E$ and
$\Psi$ corresponded to the value (the energy and frequency or the wave
function and dipole matrix element), which the correction is calculated to.
The table includes the separated contributions only to the rate. The terms
for the intensity are of the form

\[
\frac{\Delta I_{E}}{\alpha I^{(0)}} = \frac{4}{3}
\frac{\Delta\Gamma_{E}}{\alpha\Gamma^{(0)}}
\]

\noindent and

\[
\frac{\Delta I_{\Psi}}{\alpha I^{(0)}}    =
\frac{\Delta\Gamma_{\Psi}}{\alpha\Gamma^{(0)}}.
\]

One can see that the $p$-state contributions are small enough and the
estimates by means of the delta-like potential approach (see Table 1) have
to be adequate. For the decay rate, when the cancellation between $E$- and
$\Psi$-contributions is radical, and the estimates and the direct results
are in agreement within $2\delta$. The estimate by the running constant
technique (see Table 3) is in satisfactory agreement within
$1.5\,\delta_\alpha$ and the agreement is better for the higher value of
$Z$. We expect that this non-relativistic estimation is reliable within
range of the nuclear charge $Z$ up to 30.  For estimating in the both ways
of the intensity the errors are lower. Estimating according the methods
outlined above is also expected to be reliable for any transitions between
low lying levels.

The radiative correction is found to be on the one-percent level for the
nuclear charge $Z=6-10$ and it is expected to be larger for higher Z.
Thus, this effect has to be taken into consideration for interpreting of
the intensity ratio measurement in the muonic atoms.

\bigskip

\bigskip

This work has been supported in part by grant \# 95-02-03977 of the Russian
Foundation for Basic Research.

\newpage

\newcommand{\e}[1]{$\cdot10^{#1}$}
\begin{center}
\begin{tabular}{||c||c|c|c||c|c||}
\hline
\hline
&&&&&\\[-1ex]
$Z$ & $-R_\delta/\alpha$ & $\delta_1$ & $\delta_2$ &
$\Delta \Gamma_{\delta}/\alpha\Gamma^{(0)}$&
$\Delta I_{\delta}/\alpha I^{(0)}$
\\[1ex]
\hline
\hline
&&&&&\\[-1ex]
 1 & 0.10 & ~8\% & ~7\% & 0.10 & 0.24  \\[1ex]
 2 & 0.24 & 12\% & 20\% & 0.23 & 0.54  \\[1ex]
 3 & 0.34 & 11\% & 31\% & 0.33 & 0.78  \\[1ex]
 4 & 0.42 & 10\% & 40\% & 0.41 & 0.97  \\[1ex]
 5 & 0.49 & ~8\% & 46\% & 0.48 &  1.1  \\[1ex]
 6 & 0.55 & ~5\% & 52\% & 0.54 &  1.3  \\[1ex]
 7 & 0.60 & ~3\% & 56\% & 0.59 &  1.4  \\[1ex]
 8 & 0.65 & ~1\% & 59\% & 0.64 &  1.5  \\[1ex]
 9 & 0.69 & -1\% & 62\% & 0.68 &  1.6  \\[1ex]
10 & 0.73 & -2\% & 64\% & 0.72 &  1.7  \\[1ex]
\hline
\hline
\end{tabular}
\end{center}

\bigskip

Table 1. The estimates of the radiative correction arisen from the
delta-like potential calculation.

\bigskip

\begin{center}
\begin{tabular}{||c||c|c||c||}
\hline
\hline
&&&\\[-1ex]
Correction &$E$-contribution &$\Psi$-contribution  & Total\\[1ex]
\hline
\hline
&&&\\[-1ex]
$\Delta\Gamma_{\delta}/R_\delta\Gamma^{(0)}$  &
-4.00  & 3.02  & -0.98  \\[1ex]
$\Delta I_\delta/R_\delta I^{(0)}$  &
-5.33  & 3.02  & -2.32  \\[1ex]
\hline
\hline
\end{tabular}
\end{center}

\bigskip

Table 2. The main contributions to the delta-like potential calculation for
the $2p\to 1s$ transition \cite{JETP94,IK1}.

\bigskip

\begin{center}
\begin{tabular}{||c||c||c|c||}
\hline
\hline
&&&\\[-1ex]
$Z$ &
$\delta_\alpha$& $\Delta \Gamma_{\alpha}/\alpha\Gamma^{(0)}$&
$\Delta I_{\alpha}/\alpha I^{(0)}$
\\[1ex]
\hline
\hline
&&&\\[-1ex]
 3 &  100\% & 0.6 & 0.9 \\[1ex]
 4 &  ~72\% & 0.8 & 1.2 \\[1ex]
 5 &  ~58\% & 1.0 & 1.5 \\[1ex]
 6 &  ~51\% & 1.2 & 1.7 \\[1ex]
 7 &  ~46\% & 1.3 & 1.9 \\[1ex]
 8 &  ~42\% & 1.4 & 2.1 \\[1ex]
 9 &  ~39\% & 1.5 & 2.3 \\[1ex]
10 &  ~37\% & 1.6 & 2.4 \\[1ex]
15 &  ~30\% & 1.9 & 2.9 \\ [1ex]
20 &  ~27\% & 2.2 & 3.3 \\  [1ex]
25 &  ~25\% & 2.4 & 3.6 \\ [1ex]
30 &  ~23\% & 2.5 & 3.8 \\  [1ex]
\hline
\hline
\end{tabular}
\end{center}

\bigskip

Table 3. The estimates of the radiative correction inferred from the
running constant.

\bigskip

\normalsize

\begin{center}
\begin{tabular}{||c|c|c||c|c||c|c||}
\hline
\hline
&&&&&&\\[-1ex]
$Z$ & $\Delta\Gamma^s_{E}/\alpha\Gamma^{(0)}$&
$\Delta\Gamma^p_{E}/\alpha\Gamma^{(0)}$&
$\Delta\Gamma^s_{\Psi}/\alpha\Gamma^{(0)}$&
$\Delta\Gamma^p_{\Psi}/\alpha\Gamma^{(0)}$&
$\Delta\Gamma_U^{tot}/\alpha\Gamma^{(0)})$&
$\Delta I_{U}^{tot}/\alpha I^{(0)})$\\[1ex]
\hline
\hline
&&&&&&\\[-1ex]
 1 & 0.41 & -3\e{-3}   &  -0.30 & 0.01     & 0.12 & 0.26 \\[1ex]
 2 & 0.94 &   -0.02    &  -0.65 &  0.05    & 0.32 & 0.63 \\[1ex]
 3 &  1.3 &   -0.05    &  -0.90 &   0.10   & 0.51 & 0.94 \\[1ex]
 4 &  1.7 &   -0.07    &   -1.1 &   0.16   & 0.66 &  1.2 \\[1ex]
 5 &  2.0 &   -0.10    &   -1.3 &   0.21   & 0.80 &  1.4 \\[1ex]
 6 &  2.2 &   -0.13    &   -1.4 &   0.25   & 0.93 &  1.6 \\[1ex]
 7 &  2.4 &   -0.16    &   -1.5 &   0.29   &  1.0 &  1.8 \\[1ex]
 8 &  2.6 &   -0.19    &   -1.6 &   0.33   &  1.1 &  1.9 \\[1ex]
 9 &  2.8 &   -0.21    &   -1.7 &   0.36   &  1.2 &  2.1 \\[1ex]
10 &  2.9 &   -0.24    &   -1.8 &   0.40   &  1.3 &  2.2 \\[1ex]
\hline
\hline
\end{tabular}
\end{center}

\large

\bigskip

Table 4. The straightforward results for the radiative correction for the
$2p\to 1s$ transition.

\end{document}